\documentclass{nature}
\usepackage[left = 1 in, right = 1 in, top = 1 in, bottom = 1 in]{geometry}

\usepackage{xspace}     
\usepackage{textcomp}
\usepackage{graphicx}
\usepackage{caption}
\usepackage{float}
\usepackage{siunitx}
\usepackage{amsmath}
\usepackage{amsfonts}
\usepackage{amssymb}
\makeatletter
\let\saved@includegraphics\includegraphics
\AtBeginDocument{\let\includegraphics\saved@includegraphics}
\renewenvironment*{figure}{\@float{figure}}{\end@float}
\makeatother

\newcommand{\e}[1]{\ensuremath{\times 10^{#1}}}
\newcommand{\Ohm}[0]{$\Omega$\xspace}
\newcommand{\micron}[0]{\textmu m\xspace}
\newcommand{\uA}[0]{\textmu A\xspace}
\newcommand{\uW}[0]{\textmu W\xspace}

\newcommand{\Lk}[0]{\textit{L}\textsubscript{k}\xspace}
\newcommand{\Tc}[0]{\textit{T}\textsubscript{c}\xspace}
\newcommand{\VP}[0]{\textit{P}\textsubscript{V}\xspace}
\newcommand{\Rs}[0]{\textit{R}\textsubscript{s}\xspace}
\newcommand{\Vmax}[0]{\textit{V}\textsubscript{max}\xspace}
\newcommand{\Ibias}[0]{\textit{I}\textsubscript{b}\xspace}
\newcommand{\sioo}[0]{SiO\textsubscript{2}\xspace}
\newcommand{\reffig}[1]{Fig.~\ref{#1}}

\newcommand{\refcite}[1]{Ref.~\cite{#1}}
\newcommand{\about}[0]{\ensuremath{\sim}}

\newcommand{\Dc}[0]{\ensuremath{D_c}\xspace}
\newcommand{\Ic}[0]{\ensuremath{I_c}\xspace}
\newcommand{\Pc}[0]{\ensuremath{P_c}\xspace}
\newcommand{\fillfactor}[0]{\ensuremath{f}\xspace}
\newcommand{\tauon}[0]{\ensuremath{\tau}\textsubscript{on}\xspace}
\newcommand{\nwum}[0]{nW/\textmu{}m\ensuremath{^2}\xspace}
\newcommand{\fjum}[0]{fJ/\textmu{}m\ensuremath{^2}\xspace}

\providecommand{\e}[1]{\ensuremath{\times 10^{#1}}}

\begin{document}

\captionsetup[figure]{labelfont={bf},labelformat={default},labelsep=period,name={Figure}}

\title{A superconducting thermal switch with ultrahigh impedance for interfacing superconductors to semiconductors}

\author{A. N. McCaughan$^1$, V. B. Verma$^1$, S. Buckley$^1$, J. P. Allmaras$^2$, A. G. Kozorezov$^3$, A. N. Tait$^1$, S. W. Nam$^1$ \& J. M. Shainline$^1$}

\maketitle

\begin{affiliations}
 \item National Institute of Standards and Technology, Boulder, CO 80305
 \item Jet Propulsion Laboratory, California Institute of Technology, Pasadena, California 91109, USA
 \item Department of Physics, Lancaster University, Lancaster LA1 4YB, United Kingdom
\end{affiliations}

\pagenumbering{arabic} 



\begin{abstract}
A number of current approaches to quantum and neuromorphic computing use superconductors as the basis of their platform or as a measurement component, and will need to operate at cryogenic temperatures. Semiconductor systems are typically proposed as a top-level control in these architectures, with low-temperature passive components and intermediary superconducting electronics acting as the direct interface to the lowest-temperature stages. The architectures, therefore, require a low-power superconductor–semiconductor interface, which is not currently available. Here we report a superconducting switch that is capable of translating low-voltage superconducting inputs directly into semiconductor-compatible (above 1,000 mV) outputs at kelvin-scale temperatures (1\,K or 4\,K). To illustrate the capabilities in interfacing superconductors and semiconductors, we use it to drive a light-emitting diode (LED) in a photonic integrated circuit, generating photons at 1\,K from a low-voltage input and detecting them with an on-chip superconducting single-photon detector. We also characterize our device’s timing response (less than 300 ps turn-on, 15 ns turn-off), output impedance (greater than 1\,M\Ohm), and energy requirements (0.18\,\fjum, 3.24\,mV/nW).
\end{abstract}

At present, a number of quantum and neuromorphic computing architectures plan to operate at cryogenic temperatures, using superconductors as the basis of their platform~\cite{Zhang2018,King2018} or as a measurement component~\cite{Wang2018,Shainline2017,Slichter2017,Bonneau2016}. In these architectures, semiconductor systems are often proposed as a top-level control with low-temperature passive components and intermediary superconducting electronics acting as the direct interface to the lowest-temperature stages~\cite{McDermott2018}--this stratification is required because semiconductor-based amplification of small superconducting signals consumes too much power for extensive use at kelvin-scale temperatures~\cite{Patra2018,Ortlepp2013,Homulle2017}. As a result, the architectures require a low-power superconductor–semiconductor interface to, for example, leverage complementary metal–oxide–semiconductor (CMOS) coprocessors for classical control of superconducting qubits~\cite{Reilly2015}, or as a means to drive optoelectronics from superconducting detectors. However, the ability to interface superconductors with semiconductors is a missing component in these advanced computing ecosystems. 

The primary issue with interfacing superconductor electronics with semiconductor electronics is one of bandgap and impedance mismatch. The average superconductor has a bandgap almost a thousand times smaller than that of a semiconductor (e.g. 2.8 meV for Nb versus 1,100 meV for Si). Similarly, the impedances of these systems differ greatly: a typical transistor element has an effective input impedance in the 10$^4$-10$^9$ range, whereas a typical superconducting logic element will have an output impedance in the 0-10$^1$ range. Due to these mismatches, it is extremely difficult to drive the high-impedance inputs of a semiconductor element to \about1,000~mV using \about1~mV superconductor outputs. At present, there are only two known ways to generate 1,000 mV directly from a superconducting output: connect many few-millivolt devices (such as Josephson junctions) in series\cite{Benz2015}, or allow a superconducting nanowire to latch\cite{McCaughan2014}.  

The most successful previous attempts at creating a superconductor-to-semiconductor interface consist of a superconducting preamplifier stage combined with a semiconductor amplifier stage\cite{Ortlepp2013,Feng2003,VanDuzer1990}. This approach is effective at translating signal levels, but is power-constrained. In particular, using semiconductor transistors in an amplifier configuration necessarily draws significant static power (\about1~mW each), which limits scalability on a cryogenic stage. In related work, a CMOS-latch input was used after the preamplifier to limit static power\cite{Wei2011}, but this introduced the need for per-channel threshold calibration. Alternatively, it has been shown that a $>$1~V output can be created from a nanowire device such as the nanocryotron\cite{McCaughan2014}, but using the nanocryotron as a means for semiconductor-logic interfacing has drawbacks: creation of the high-impedance state is a relatively slow hotspot-growth process along the length of the nanowire (0.25\,nm/ps in NbN\cite{Berggren2018}); it is hysteretic and not able to self-reset without external circuitry; and output-input feedback is a concern, as the input and output terminals are galvanically connected\cite{Zhao2017}.

\begin{figure} 
\centering
    \includegraphics[width=6.5in]{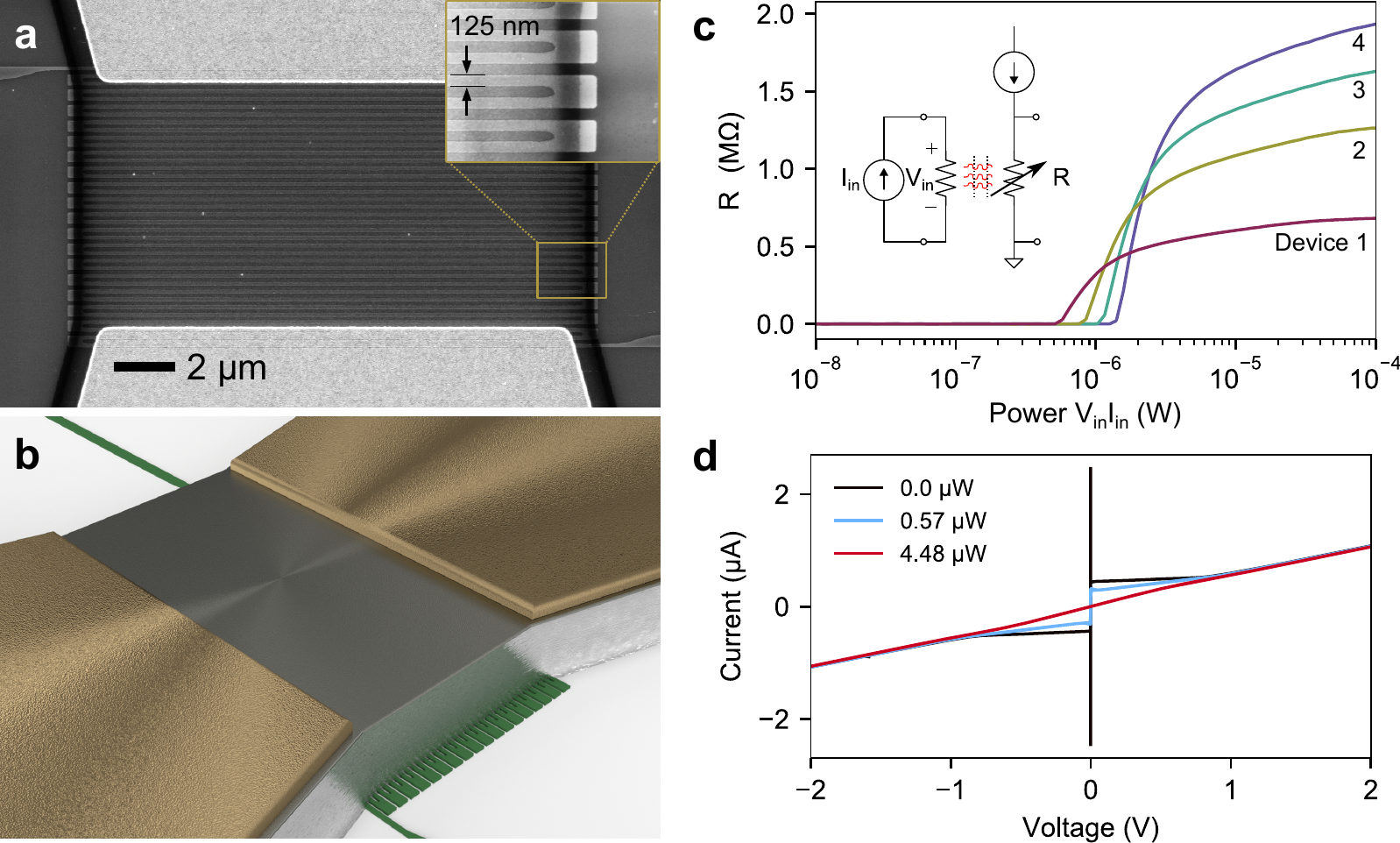}
    \caption{High-impedance superconducting switch overview. (a) Scanning electron micrograph of one device (inset) closeup of the nanowire meander. (b) Schematic illustration of the device, showing the three primary layers (resistor, dielectric, and nanowire) as well as contact pad geometry. (c) Resistance data versus input power for several devices and circuit schematic for resistance measurement. Maximum resistance is proportional to device area, with devices 1-4 having areas 44, 68, 92, and 116~\textmu{}m$^2$. (d) I-V curve of one device for three different input powers.}
\label{overview}
\end{figure}

In this Article, we report a monolithic switch device that can translate low-voltage superconducting inputs directly into semiconductor-compatible ($>$1,000 mV) outputs. The switch combines a low-impedance resistor input (1-50 \Ohm) with a high-impedance ($>$1~M\Ohm) superconducting nanowire-meander switch element. The input element and switching element are isolated galvanically but coupled thermally by a thin dielectric spacer (25~nm \sioo). When input current is applied to the resistor, the state of the entire nanowire-meander is switched from superconducting to normal. The input induces an extremely large impedance change in the output: from 0\,\Ohm to $>$1\,M\Ohm (\reffig{overview}). The power cost of inducing this change is surprisingly small when compared to existing methods, and crucially it can be operated in a non-hysteric (that is, self-resetting) regime. As a demonstration of a superconductor-semiconductor interface, we have used the switch to drive an LED in a photonic integrated circuit, generating photons at 1~K from a low-voltage input and detecting them with an on-chip superconducting single-photon detector.


\section*{High-impedance superconducting switch}

Our device consists of a 3-layer stack (\reffig{overview}b). On the top of the stack is a resistor made from a thin film of normal metal with a small resistance (1-50\,\Ohm). On the bottom of the stack is a meandered nanowire patterned from a superconducting thin film. The nanowire layer acts as a high-impedance, phonon-sensitive switch, while the resistor layer is used to convert electrical energy into Cooper-pair-breaking phonons. Like related low-impedance thermal devices\cite{Lee2003,Zhao2018}, between these two layers is a dielectric thermal spacer that has two purposes: thermally coupling the resistor layer to the nanowire layer, and electrically disconnecting the input (resistor) from the output (nanowire switch). The device has four terminals total, with two of the terminals connected to the resistor and two of the terminals connected to the nanowire (\reffig{overview}). Fabrication details are available in the Methods section.

The device begins in the ``off'' state where there is no electrical input to the resistor and the nanowire has a small current-bias. To transition to the ``on'' state, a voltage or current is applied to the input terminals of the resistor and thermal phonons are generated. The thin dielectric carries phonons generated from the resistor to the nanowire. Phonons with energy $>$2$\Delta$ break Cooper pairs within the nanowire, destroying the superconducting state of the nanowire. Once the superconducting state has been completely destroyed in the entire nanowire, the device is in the on state. 

Analogously, this process can be described in terms of an effective temperature: the dielectric layer is thin enough that the phonon systems between the nanowire and the resistor are tightly coupled, meaning the phonon temperature in the nanowire is closely tied to the temperature of the resistor. When enough electrical power is delivered to the resistor, the nanowire is driven above its critical temperature and becomes normal, jumping from 0\,\Ohm to $>$1\,M\Ohm. Once the device has switched, current is then driven into the high-impedance output load and a large voltage can be generated.

When characterizing a switch, of primary importance is its on and off resistance. We measured the steady-state behavior of the switch by applying power to the resistor inputs of several devices and measuring the nanowire resistance with an AC resistance bridge. As can be seen in  \reffig{overview}, each device remained at zero resistance until a critical surface power \Pc was reached. When more than \Pc was applied to the resistor, the resistance of the underlying nanowire increased rapidly, ultimately saturating at the normal-state resistance of the device. One potential concern was that phonons from the resistor could escape in-plane (e.g. out through the thick gold leads), resulting in wasted power. This type of edge power loss would scale with the length of the edge, and so we measured \Pc for devices of several sizes.  However, by dividing each device's  \Pc by its active area $A$, we found that the devices had critical surface power densities $\Dc = \Pc/A$ of 21.0\,$\pm$\,0.6\,\nwum.  This means power loss through edge effects (e.g. along the substrate plane or into the gold contacts) did not play a role at the scale of device measured here. Additionally, through thermal modeling, we also found that worst-case thermal crosstalk between adjacent devices would be negligible if the devices were separated by a few micrometers (further details on lateral heat transport and crosstalk are available in the Supplementary Information). As shown in \reffig{overview}c, the resistance of each device continues to increase beyond $\Pc$ -- this was likely due to non-uniform dissipation within the resistor element creating local temperature variation.  Note that performing this measurement with a low-power measurement technique (such as an AC resistance bridge) was critical in order to limit Joule heating from current passing through the nanowire element. In this experiment, we applied a maximum of 10 nA (\about100 pW) in order to guarantee that the nanowire was not heated by the measurement process. The 4.5-nm-thick tungsten silicide (WSi) film used for the nanowires had a \Tc of 3.4~K, and all measurements were taken at a base temperature of 0.86~K.

\section*{Driving a cryogenic LED}

\begin{figure}[H] 
\centering
    \includegraphics[width=6.5in]{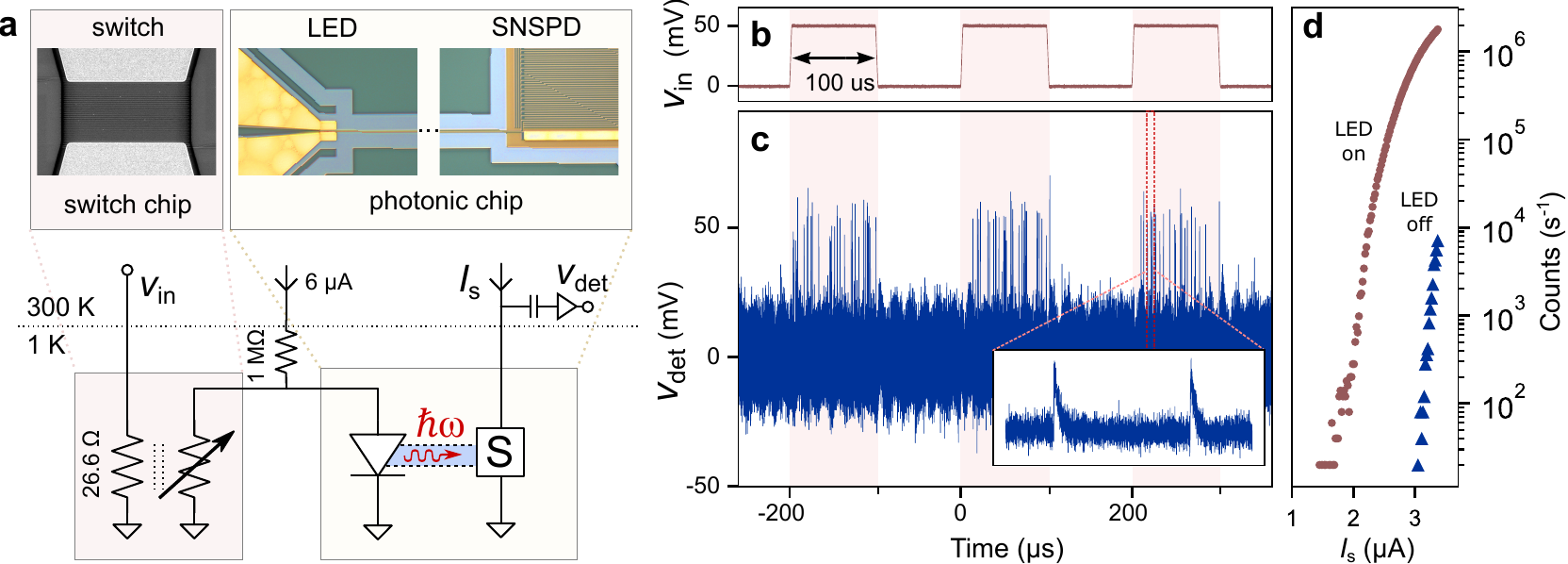}
    \caption{Driving a photonic integrated circuit at 1~K. (a) Schematic and circuit setup for powering a cryogenic LED with the switch and reading out the generated light using a waveguide-coupled single-photon detector. (b,c) Switch input and detector output versus time. When $v_{in}$ is high, photons generated by the LED are transmitted via waveguide to a superconducting nanowire single-photon detector, producing detection pulses. (inset) Zoom-in of the detector output pulses. (d) Detector count rates for the experiment with the LED on (red) and off (blue).}
\label{led}
\end{figure}

As a means of demonstrating the superconductor-semiconductor interface, we used the switch to dynamically enable a cryogenic LED in a photonic integrated circuit (PIC) using only a low-level input voltage. Shown in \reffig{led}, the output from the switch was wirebonded to a PIC that had an LED which was waveguide-coupled to a superconducting nanowire single-photon detector (SNSPD). The switch translated the 50~mV input (\reffig{led}b) signal into 1.12~V at the output, enabling and disabling the LED in a free-running mode. Photons produced by the LED were coupled via waveguide into the detector, producing clicks on the detector output (\reffig{led}c). The switch was driven with 94.0~\uW of input power (55.9~\nwum, well above \Dc), generating an on-state resistance of approximately 400~k\Ohm. We note these particular LEDs had a low overall efficiency (\about$10^{-6}$ as characterized in \refcite{Buckley2017}), and so a large LED input power was necessary to generate the handful of photons per period. The large LED power requirement necessitated wide nanowires (1~\micron wide for this experiment) to carry the requisite current, and so the device area and input power scaled proportionally.  Steady-state behavior of the circuit is shown in \reffig{led}d, which characterizes the detector response with the switch input power above and below the \Dc threshold. We additionally verified that the counts measured on the detector were in fact photons generated by the LED--not false counts due to sample heating or other spurious effects--by reducing the LED bias below threshold and observing no clicks on the detector output, and also by quadrupling the switching input power and observing no heating-induced change in count rate.

\section*{Transient and sub-threshold response}

\begin{figure}[H] 
\centering
    \includegraphics[]{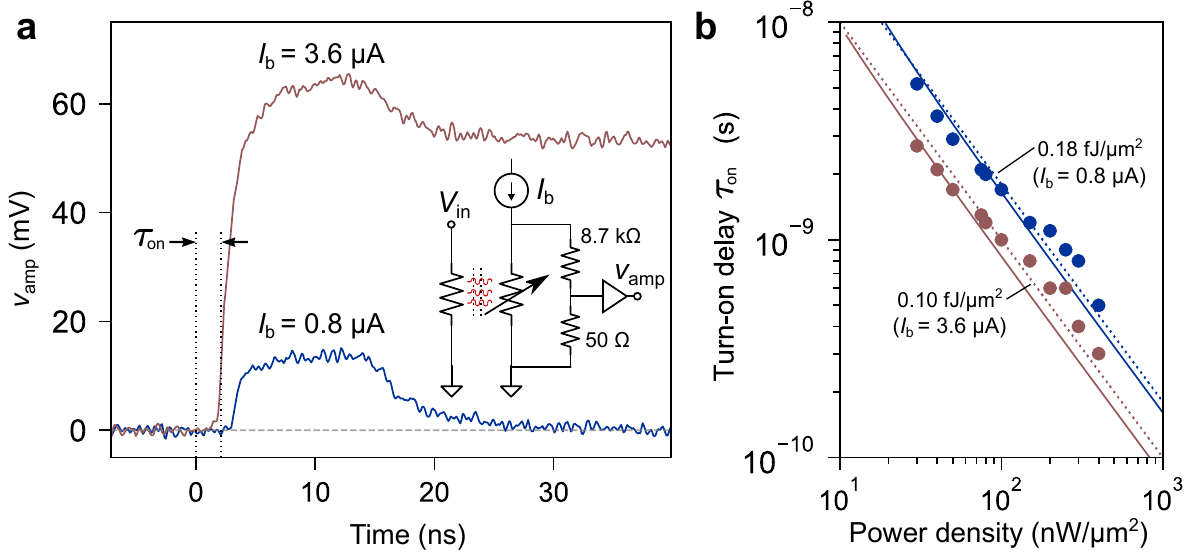}
    \caption{Driving an 8.7 k\Ohm load using the switch. (a) Output produced by a 10-ns-wide square pulse to the heater with input surface power density $D = 50$~\nwum, highlighting the latching and non-latching regimes. Trace data taken with a 1~GHz bandwidth-limited amplifiers and oscilloscope. (b) Turn-on delay \tauon versus applied input power when the nanowire is biased below the retrapping current (blue) and near \Ic (red). The solid lines are fits generated by the ballistic phonon transport modeling, and the dotted lines are constant-energy per unit area curves corresponding to 0.18~\fjum (blue) and 0.10~\fjum (red).}
\label{step}
\end{figure}

We also characterized the transient properties of the device when driving high-impedance loads by placing a 8.7~k\Ohm on-chip resistor at the output of the device. In this experiment, we applied voltage pulses to the resistor input with a pulse generator and measured the device output, while applying a current bias to the nanowire either below the retrapping current (\reffig{step}, red data), or near the critical current (\reffig{step}, cyan data). As seen in the circuit diagram of \reffig{step}, an output voltage could only be generated when the switch reached a significant resistance ($\gg$1k\Ohm) allowing us to probe the impedance transition of the device. The results from this experiment, shown in \reffig{step}, showed that the device could turn-on from its low-impedance state to its high-impedance state below 300~ps, characterized by a power-delay product on the order of \about100 aJ per square micrometer of device area. 

Crucially, the impulse-response also demonstrates that this device can self-reset. As highlighted in \reffig{step}a, when the nanowire is biased below the retrapping current, it becomes non-hysteretic and returns to the zero-voltage (superconducting) state after the input is turned off.  The logic follows simply: below the retrapping current, the self-heating caused by the nanowire bias current does not generate enough power to keep the wire above \Tc. Thus, when the additional heating from the resistor is removed, the nanowire is forced back into the superconducting state--it does not get stuck in the on-state or ``latch.'' This non-latching property is in contrast to existing thin-film nanowire devices previously developed, and is critical to guarantee device reset in unclocked systems where the input bias to the devices is not periodically turned off. We note that even in this regime, there is still a thermal recovery time constant for the device to transition from the normal (on) to superconducting (off) state. We found that fall time was on the order 10\,ns, which is consistent with the thermal recovery time constants previously reported for WSi. Additionally, it should be noted that although the nanowire fabricated here has a very high kinetic inductance, the L/R time constant of the switch--that could potentially limit the rise time of the current output--is not a limiting factor. When the device transitions from low- to high-impedance, the expected \Lk/\Rs is equal to 0.39 ps. 

\begin{figure} 
\centering
    \includegraphics[]{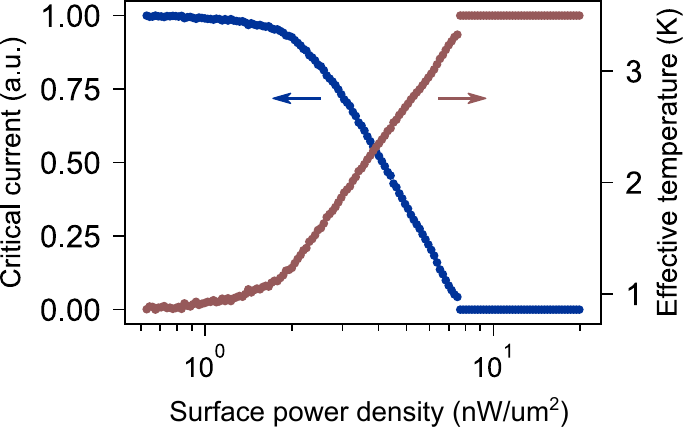}
    \caption{Critical current and inferred temperature versus input power density. As the input power density is increased, the effective temperature of the meander rises, and its critical current is reduced. The small discontinuity is an experimental artifact from measuring very small critical current values.}
\label{ic}
\end{figure}

To better understand the response of the device below the critical surface power density \Dc, we measured the nanowire critical current as a function of power applied to the resistor.  The result of this characterization is shown in \reffig{ic}. For each datapoint, we applied a fixed amount of electrical power to the input resistor and measured the critical current of the nanowire several hundred times, taking the median value as $I_c(P)$. We then extracted an effective temperature for the nanowire by numerically inverting the Ginzburg-Landau relation $I_c(t) = I_{c0} (1-t^2)^{3/2} (1+t^2)^{1/2}$ where $I_{c0}$ was the critical current at zero applied power and $t$ was the normalized temperature of the device $T/T_c$ (for this material, the $T_c$ was measured to be 3.4~K). Note the non-uniform heating causes the critical current to reach zero at 8\,\nwum--well before the jump in resistivity shown in \reffig{overview}c at \Dc (21\,\nwum)--because localized heat can suppress \Ic but the resistivity measurement accounts for the state of every part of the nanowire. The data in \reffig{ic} show that there is a non-linear relationship between the nanowire temperature and the applied heating power. We note that these nanowires are likely constricted due to current crowding\cite{Clem2011a} at the bends, and as a result may obfuscate changes in \Ic at low powers in \reffig{ic}. 

\section*{Thermal transport modeling}

A complete model of the dynamics of the device requires an investigation of the nonequilibrium dynamics of the electron and phonon systems of the heater, dielectric spacer, and nanowires. While a full description is beyond the scope of this paper, we find that an approximation using ballistic phonon transport is sufficient for describing the main experimental results.  Given that the bulk mean-free-path of phonons in \(\text{SiO}_2\) is on the order of \SI{1}{\micro\meter} at \SI{2.5}{\kelvin}, we assume that phonons escaping from the heater travel through the dielectric without scattering and either interact with the nanowire or continue unimpeded to the substrate\cite{Allmaras2018}.  Within this model and under the simplifying assumption of equilibrated electron and phonon systems in the nanowire at temperature \(T_{WSi}\), the energy balance equation of the nanowire is given by
\begin{equation}
\left(C_{e}\left(T_{WSi}\right) + C_{ph}\left(T_{WSi}\right)\right)d\:\fillfactor \frac{\partial T_{WSi}}{\partial t} = \fillfactor  \chi_{abs} P_{h} - \Sigma \left({T_{WSi}}^4 - {T_{sub}}^4\right)
\end{equation}
where \(C_e\left(T_{WSi}\right)\) is the BCS electron heat capacity, \(C_{ph}\left(T_{WSi}\right)\) is the lattice heat capacity, \(d\) is the nanowire thickness, \fillfactor is the nanowire fill factor, \(\chi_{abs}\) is the fraction of energy incident on the nanowire which is absorbed, \(\Sigma\) describes the magnitude of phonon energy flux from the nanowire to the substrate per unit area, and \(P_{h}\) is the power dissipated by the heater per unit area. For a WSi device with \(d =\) \SI{4.5}{\nano\meter}, \fillfactor = 0.5, sheet resistance \(\rho_{sq} = \) \SI{590}{\ohm/sq}, \(T_c = \) \SI{3.4}{\kelvin}, and using values from the literature \cite{Sidorova2018}\cite{Marsili2016} for the diffusion coefficient \(D = \) \SI{0.74}{\centi\meter^2/\second}  and specific heat ratio \(C_e\left(T_c\right)/C_{ph}\left(T_c\right) \sim 1\), the parameters \(\chi_{abs}\) and \(\Sigma\) are chosen to fit the turn-on delay vs dissipated power results of~\reffig{step}b.  The calculated curves show the turn-on delay for temperature thresholds of \SI{2.5}{\kelvin} and \SI{3}{\kelvin}, where temperature threshold refers to the minimum temperature required at a given bias current to switch the device. The two remaining free parameters were fitted to be \(\chi_{abs} = 0.02\), and \(\Sigma = \) \SI{0.7}{\watt/\meter^2 \kelvin^4}. This value of \(\Sigma\) corresponds to nonbolometric phonon bottle-necking at the \(\text{WSi/SiO}_2\) interface with a conversion time from the non-escaping to escaping group of phonons with a magnitude over \SI{1}{\nano\second}, which is consistent with experiment \cite{Sidorova2018}.  For comparison, the estimation of the parameter \(\chi_{abs}\) based on the solution of ballistic phonon transport in the nanowire is provided in the Supplementary Information. 

While the basic principle of operation is applicable to materials with higher critical temperatures, the details of the heat transfer between heater and superconductor would change.  The heat capacity of materials increases with temperature, so more energy will be required to heat the superconductor to its current dependent critical temperature.  With a larger superconducting gap \(\Delta\), higher-energy phonons are required to break Cooper pairs.  At the same time, higher temperature operation means that phonons from the heater will have higher energies, and shorter mean-free-paths in the dielectric, which will lead to additional scattering and absorption.  It is currently unclear if this will make heat transfer less efficient due to the additional scattering, or more efficient by keeping energy trapped in the local area of the nanowire.  

A useful figure of merit for these devices is \VP, the output voltage generated per unit input power while the switch is on. Due to proportionalities between the device resistance and area, and also between nanowire width and \Ic, this figure of merit is area- and shape-independent--it depends only on the materials used and the nanowire configuration. We calculated \VP by first noting that the power required to heat the full area $A$ of a given device was $\Dc A$. For a device with nanowires of width $w$, thickness $t$, and fill-factor $f$, the amount of switching resistance generated in that area is $R_s f A/w^2$, where $R_s$ is the nanowire normal-state sheet resistance. For the WSi material used here, the bias current density $J$ was 1.6\e{9}~A/m$^2$ (non-latching) or 7.2\e{9}~A/m$^2$ (latching), and $R_s$ was 590~\Ohm/sq. The resulting voltage generated per unit power is then $R_s f J/w^2 \Dc$ (thus, independent of device area), and so for the devices characterized in \reffig{overview}, \VP was 0.72~mV/nW (non-latching) or 3.24~mV/nW (latching).

One potential area of concern when using this device as a switch is the power usage from current-biasing the device. Typically, a current-bias capable of driving high voltages require a large amount static energy dissipation: when using a resistor or MOSFET-based current source as a current bias, to achieve a maximum voltage of \Vmax will generally require $\Ibias\Vmax$ of static power. However, for these devices a better approach will be to use inductive biasing to generate the high-impedance current bias. Superconducting thin films such as the one used here lend themselves particularly well to the generation of large inductances in compact areas, due to their large kinetic inductance. For instance, to generate a 200~mV swing on a CMOS input capacitance of 5~fF requires a bias current of 50~\uA being carried by an 160~nH inductor -- a trivial amount of inductance to generate with a superconducting nanowire. More importantly, the inductor can be charged only when needed, using a low-voltage superconducting element such as a Josephson junction.

\section*{Conclusions}

Our superconducting thermal switch has a number of favourable features as a communications device between superconducting and semiconducting elements: it provides switch impedances of more than 1M\Ohm, input–output isolation, low turn-on time, and low-power operation with zero passive power required. Even with the reset-time limitations presented by thermal recovery, this switch has particular applicability for driving optoelectronics on a cryogenic stage such as LEDs or modulators. In applications such as quantum photonic feed-forward experiments or low-power neuromorphic hardware\cite{Shainline2018e}, clicks from efficient superconducting detectors need to be converted to optoelectronic-compatible signals, and a nanosecond-scale thermal recovery is acceptable as long as the initial response is fast. We note for these types of applications, it may be best to configure the device geometry to minimize propagation delay -- the propagation delay for powering an LED to 1~V will be much smaller if 1~mA is driven across a switch of 1~k\Ohm, rather than driving 1~\uA across 1~M\Ohm. It may also be helpful to drive the device with another superconducting three-terminal device\cite{McCaughan2016a}, as this can provide a purely non-resistive superconducting input and be fabricated in the same step as the nanowire meander.

Looking forward, there are a number of practical methods to enhance the operation of this device, depending on what tradeoffs are acceptable in a given application. The simplest would be to use multiple layers of nanowire: the on-resistance could, for example, be effectively doubled by adding an additional nanowire meander underneath the first (at some minor turn-on energy cost). If power usage is a concern, the on-state power requirements could be decreased by placing the device on a membrane. This would greatly increase the thermal resistance, reducing overall energy cost at the cost of increasing the thermal turn-off time. For higher operational frequencies, a different nanowire material could be used (for example, NbN for a \about1~ns thermal reset time~\cite{Kerman2009}). Finally, this device does not fundamentally need to be a thermal device: any method of inducing a phase change in a superconducting film -- for instance, using an electric-field induced superconductor-to-insulator
transition~\cite{Ueno2008} -- could be operated equivalently.

\section*{Methods}

\subsection*{Fabrication details}

Fabrication began with a clean thermal oxide wafer (150~nm \sioo on Si). WSi was sputtered uniformly over the entire wafer to a thickness of 4.5~nm, and afterwards--but before breaking vacuum--a thin capping layer (1-2~nm) of amorphous Si was also sputtered on top. (WSi was chosen primarily for its high practical fabrication yield in our lab--other highly-resistive thin-film superconductors should work equivalently, although with potential power/thermal tradeoffs discussed in the thermal modeling section.)  Next, contact pads for the superconducting layer were patterned using a liftoff process and deposited by evaporating 5~nm~Ti / 100~nm~Au / 5~nm~Ti.  We then patterned and etched the WSi layer to form the nanowires. Afterwards, we sputtered the whole wafer with 25~nm of \sioo.  \sioo was chosen because of its compatibility with WSi -- in past experiments we had found that \sioo deposition did not negatively impact the superconducting parameters of the WSi layer (\Tc, \Ic, etc.). Using a liftoff process, we then fabricated the resistor layer by evaporating 15~nm of PdAu.  Lastly, the low-resistance contact pads were deposited using another liftoff process, evaporating 5~nm Ti followed by 100~nm of Au.

\begin{addendum}
 \item[Acknowledgements] The authors would like to thank Florent Lecocq for helpful discussions, and Adriana Lita for insight into the fabrication development. The U.S. Government is authorized to reproduce and distribute reprints for governmental purposes notwithstanding any copyright annotation thereon. Part of this research was performed at the Jet Propulsion Laboratory, California Institute of Technology, under contract with the National Aeronautics and Space Administration.  J.P.A. was supported by a NASA Space Technology Research Fellowship.  Support for this work was provided in part by the DARPA Defense Sciences Offices, through the DETECT program.  
 \item[Author contributions] A.N.M., V.V., S.M.B., and J.M.S. conceived and designed the experiments. A.N.M. performed the experiments. J.P.A. and A.G.K. analyzed and modeled the thermal properties of the device. A.N.M. and V.V. fabricated the devices. A.N.M., A.T, and S.W.N. analyzed the data.
 \item[Competing Interests] The authors declare U.S. Patent US10236433B1
 \item[Data availability] The data that support the findings of this study are available within the paper. Additional data are available from the corresponding authors upon reasonable request.
\end{addendum}

\newpage

\bibliographystyle{naturemag}

\newpage

\section*{Supplementary Information}

\section*{Thermal transport modeling - estimation of \(\mathbf{\chi_{abs}}\)}

We estimate the parameter \(\chi_{abs}\) based on ideal ballistic propagation of phonons from the heater to the nanowire through the thin dielectric.  To simplify the analysis, we assume that the heater instantaneously reaches the stationary response due to the dissipated power at the time when an electrical pulse is applied.  Phonons in the WSi travel along ballistic trajectories with the average sound velocity \(v\) with an angle \(\theta\) with respect to normal of the interface and we consider only the absorption of phonons, neglecting re-emission in this treatment.  The electron system of the nanowire is assumed to have an equilibrium distribution described by the temperature \(T_e\).  Under these assumptions, the kinetic equation describing the phonon distribution in the nanowire \(N_{WSi}(\omega, z, \theta, t)\) with depth in the \(z\) direction from \(0 \le z \le d\) is given by
\begin{equation}
\frac{\partial N_{WSi}}{\partial t} + v \cos(\theta) \frac{\partial N_{WSi}}{\partial z} = -\frac{N_{WSi} - N^0\left(T_e\right)}{\tau_{ph-e}}
\end{equation}
where \(N^0\left(T_e\right)\) is the Planck distribution at the electron temperature of the nanowire and \(\tau_{ph-e}\) is the phonon-electron interaction time. Under the assumption of a stationary input phonon distribution \(N^H\left(T_H\right)\) from the heater at \(z = 0\), the solution to this partial differential equation becomes
\begin{equation}
N_{WSi}\left(\omega, z, \theta \right) = N^0(T_e) + \left[N^H(T_H)- N^0(T_e)\right] e^{-\frac{z}{l_{ph-e} \cos(\theta)}}
\end{equation}
where \(l_{ph-e} = v\tau_{ph-e}\) is the energy-dependent phonon-electron interaction length.  The energy transferred to the WSi is expressed as \(Q_{WSi}^{in} - Q_{WSi}^{out} = P \chi_{abs} \fillfactor = Q_{WSi}^{in} \chi_{abs}\). The energy flux is evaluated according to
\begin{equation}
Q(z) = \int_0^{\omega_D}d\omega \rho(\omega) \hbar \omega v\int_0^{\theta_m} d\theta \sin(\theta)\cos(\theta)N(\omega, z, \theta)
\end{equation}
where \(\rho(\omega)\) is the phonon density of states and \(\omega_D\) is the Debye frequency.  Using this form, we evaluate \(\chi_{abs}\) by rearranging terms and evaluating \(Q_{WSi}^{in}\) at \(z = 0\) and \(Q_{WSi}^{out}\) at \(z = d\), leading to 
\begin{equation}
\chi_{abs} = \frac{\int_0^{\omega_D}d\omega \rho(\omega) \hbar \omega v\int_0^{\theta_m} d\theta \sin(\theta)\cos(\theta) \left[N^H(T_H)- N^0(T_e)\right]\left(1 - e^{-\frac{d}{l_{ph-e} \cos(\theta)}}\right)}{\int_0^{\omega_D}d\omega \rho(\omega) \hbar \omega v\int_0^{\theta_m} d\theta \sin(\theta)\cos(\theta)N^H(T_H)}
\label{Eq: Chi Absorption Full}
\end{equation}
We take a series expansion of the exponential function because \(\frac{d}{l_{ph-e} \cos(\theta)} \ll 1\) for thin WSi nanowires.  The interaction length is given by \(\frac{1}{l_{ph-e}(\omega)} \approx \frac{\gamma}{v \tau_0}\left(\frac{\hbar \omega}{k_B T_c}\right) \) for \(\hbar \omega \geq 2\Delta\) and 0 otherwise.  This occurs because only phonons with energies greater than \(2\Delta\) are capable of breaking Cooper pairs.  The parameter \(\gamma = \left.\frac{8 \pi^2}{5}\frac{C_e}{C_{ph}}\right\rvert_{T_c}\) describes the ratio of electron to lattice heat capacity at \(T_c\). We can neglect the \(N^0(T_e)\) term in (\ref{Eq: Chi Absorption Full}) because the WSi temperature is significantly less than the heater temperature for the power dissipation levels measured experimentally, and we confirm that this approximation is valid by numerical calculation of the full and approximate expressions. Under these simplifications, and assuming low temperature, we arrive at the expression
\begin{equation}
\chi_{abs} = \frac{1}{\cos\left(\frac{\theta_m}{2}\right)^2}\frac{\gamma d}{v \tau_0}\left(\frac{T_H}{T_c}\right)\frac{\int_{2 \Delta/k_B T_H}^{\infty}dx\:  \frac{x^4}{e^{x} - 1}}{\int_0^{\infty}dx \frac{x^3}{e^{x} - 1}} \text{.}
\label{Eq: Chi Abs Simple}
\end{equation}
To determine \(\chi_{abs}\), we estimate \(\tau_0 =\) \SI{5000}{\pico\second} based on measurements\cite{Sidorova2018} of \(\tau_{ep}\) and the parameters listed in the main text.  The heater temperature is determined using thermodynamic properties of Au and Pd and an estimated phonon escape time of \SI{100}{\pico\second}.

\begin{figure}[H] 
\centering
\includegraphics[]{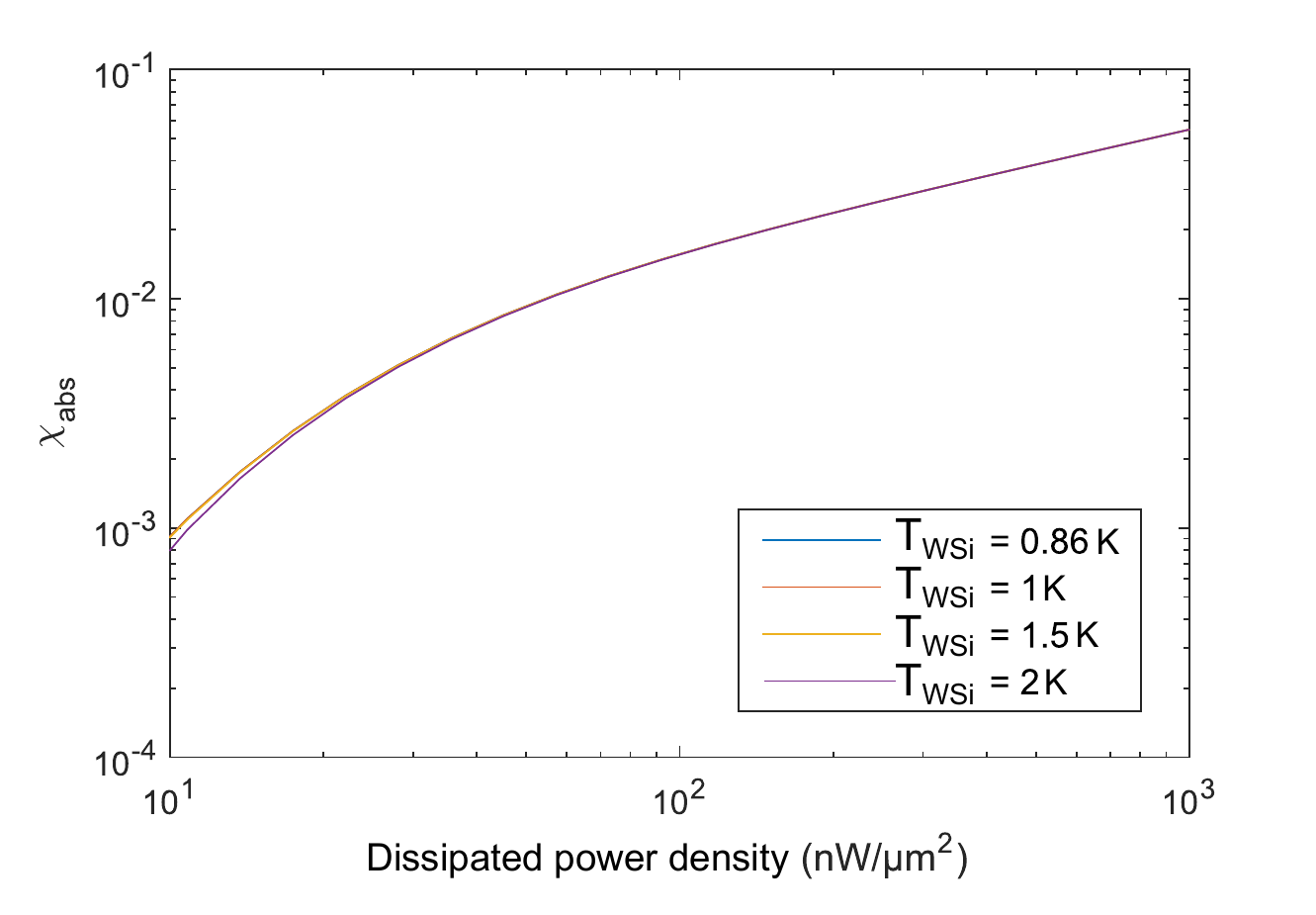}
\caption{Estimation of \(\chi_{abs}\) as a function of input power density. The value of \(\chi_{abs}\) has a weak dependence on temperature in the range of interest.}
\label{X Estimate}
\end{figure}
This calculation leads to a value of \(\chi_{abs}\) which is between 0.01 and 0.03 over the dissipated power range measured in the turn-on delay experiment, which is the same range needed to match experiment based on the simple fixed \(\chi_{abs}\) model.  The absorption fraction increases as the dissipated power increases due to the increased number of high energy phonons present in the heater radiation.  These high energy phonons have a shorter \(l_{ph-e}(\omega)\), leading to a higher fraction of energy absorbed in the nanowire.  However, this simplified model neglects the influence of \(\text{SiO}_2\) scattering, which also increases as phonon energy increases and is expected to limit the magnitude of this change in \(\chi_{abs}\).  Consequently, the ballistic propagation assumption used in this model is violated for these higher-energy phonons.  A fully quantitatively accurate model of phonon transport must also consider the reflection of phonons off both the \(\text{SiO}_2\)/WSi and \(\text{SiO}_2\)/Si interfaces.  While transmission from \(\text{SiO}_2\) to WSi is estimated to be approximately 70\% based on the acoustic mismatch model for most incident angles, total internal reflection of acoustic modes is predicted at the \(\text{SiO}_2\)/Si interface for incidence angles greater than 50 degrees, which will increase the number of phonons available for absorption in the nanowire layer.

\section*{Crosstalk and lateral heat transport}

\begin{figure} 
\centering
    \includegraphics[]{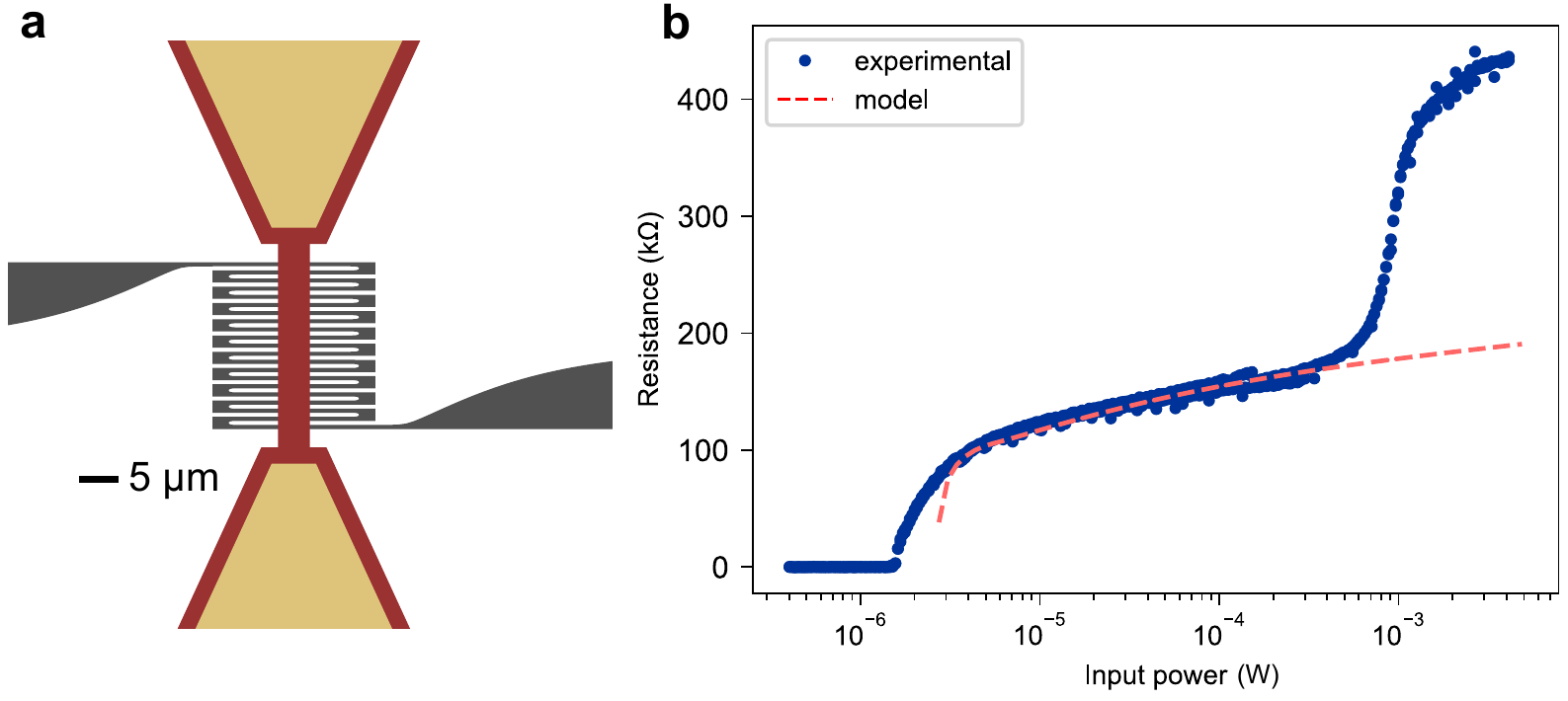}
    \caption{Design and results of edge-leakage test. (a) Geometry of the edge-leakage test, showing a narrow resistor element (red) routed through the middle of a large nanowire meander (black). (b) Experimental data and theoretical modeling of the device resistivity as a function of temperature for the edge-leakage test. Adding a temperature dependence of the resistance versus temperature curve would smooth the onset of finite resistance at low heater power.}
\label{S1}
\end{figure}

In order to determine whether the device had significant in-plane heating, we designed a device with a narrow heater-resistor centered in a nanowire meander.  The geometry can be seen in \reffig{S1}.  In this experiment, we applied power to the resistor and measured the resulting resistance of the nanowire meander. The results, shown in \reffig{S1}b, indicate that there is very little in-plane heating: at powers around \Dc, we see a resistance in the nanowire form that is equivalent to the superconducting material directly underneath the resistor becoming normal. Only at much greater applied powers (\about1~mW) does this resistance increase significantly indicating that a much greater amount of power is necessary to heat the superconducting material in the neighborhood of the resistor--likely due to heating the entire device substrate, as was observed in \refcite{Allmaras2018}. 

\begin{figure}[H] 
\centering
    \includegraphics[width=3in]{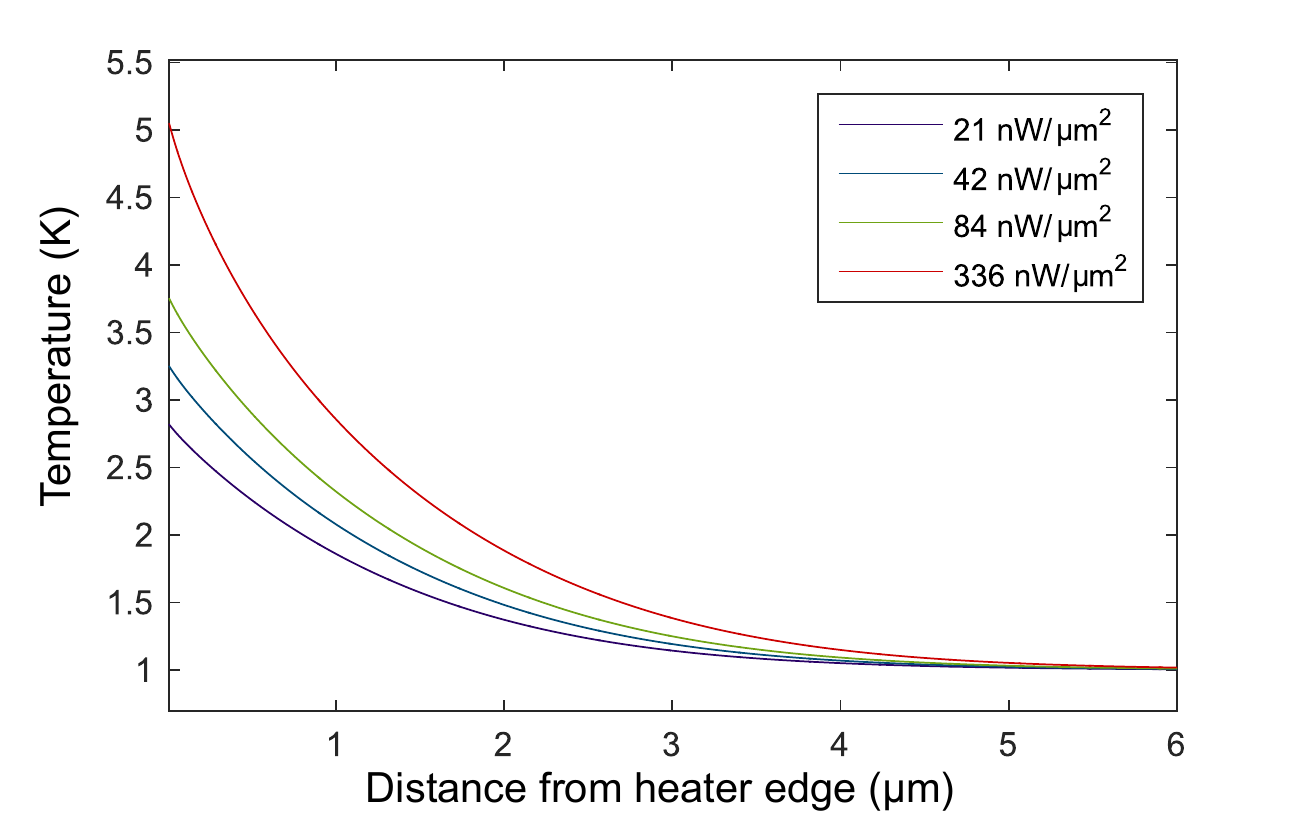}
    \caption{Distance from the edge of the heater at which the dielectric reaches a given temperature as a function of heater power density.}
    \label{Lateral Heat Transport}
\end{figure}

The experimental results are consistent with thermal modeling of the lateral heat transport in thin \(\text{SiO}_2\).  By using the approach of \refcite{Allmaras2018} with the same form and value of the thermal coupling between the \(\text{SiO}_2\) and Si for a bath temperature of \SI{1}{\kelvin}, we calculated the temperature of the dielectric layer as a function of distance from the edge of the heater using a worst-case scenario where the device was always being heated.  The resulst are shown in \reffig{Lateral Heat Transport}.  We found that the surrounding temperature--and thus thermal crosstalk--falls off within a distance of a few micrometers. The upper limit of scalability ultimately depends on the thermal coupling from the substrate to the cold bath and the cooling power of the fridge, but for this particular material stack, the devices could be patterned quite densely.

\end{document}